# Realization of Friedrich–Wintgen QBIC with high Q-factors based on acoustic–solid coupling and sensing application

Bowei Wu [#] , BoyueSu [#] , Shuanghuizhi Li, Tingfeng Ma [*]

Zhejiang-Italy Joint Lab for Smart Materials and Advanced Structures, School of Mechanical Engineering and Mechanics, Ningbo University, Ningbo 315211, China

## Abstract

In recent years, bound states in the continuum (BICs) have attracted extensive attentions in the sensing field due to their theoretically ultra-high resonance quality factors (Q-factors). Among them, Friedrich–Wintgen (F–W) BICs, which arise from the interference between different coupled modes, are particularly promising for acoustic sensing applications owing to the easy realization. Most existing F–W BICs are realized in open systems through the interference between waveguides and resonant cavities. However, with increasing demands for higher resolution and sensitivity in modern chemical and biological sensing, the practically measured Q-factors of conventional open-system F–W BICs often fall short of expectations.

In this work, we introduce Fabry–Pérot (F–P) resonance via acoustic–solid coupling to explore the formation mechanism and realization method of high-Q F–W BICs in quasi-closed systems, and further investigate their application in gas sensing. A coupled resonator model combining elastic and acoustic waves in a quasi-closed cavity is first established. Coupled mode theory is employed to calculate the eigenmodes of both localized and radiative modes. Based on this, the Hamiltonian matrix of the coupled system is constructed, from which the acoustic transmission spectrum is derived. The results show that the Q-factor of the F–W BIC induced by acoustic–solid coupling is significantly higher than that of open systems, which is further validated by experiments.

Based on this, a gas concentration sensing technique based on acoustic–solid coupled F–W BIC behavior is developed. A sensing device is fabricated accordingly, and gas concentration measurements are carried out. Experimental results demonstrate a pronounced response to gases with different concentrations, confirming the feasibility and reliability of this novel gas sensing approach.

* matingfeng@nbu.edu.cn

# 1. Introduction

Acoustic sensing has demonstrated expansive applications potential in environmental monitoring, industrial inspection, and biomedical diagnostics, owing to its non-contact nature, high sensitivity, and adaptability to complex environments. In particular, in fields such as gas detection and sound source localization, acoustic sensing has become a research focus due to its sensitive response to sound pressure, frequency, and propagation characteristics. In recent years, extensive studies have been conducted on various types of acoustic sensors, including phononic crystal structures[1-5]. Numerous innovative designs and optimization strategies have been proposed, continuously advancing acoustic sensing technologies toward higher sensitivity and multi-functionality. For example, Chen[6] proposed a gradient phononic crystal structure in which topological edge states are introduced by breaking spatial inversion symmetry. By tuning the dispersion relation through parameter gradients, an acoustic rainbow trapping effect was achieved. Both numerical and experimental results demonstrated its robustness and enhanced sensing performance, offering a new approach to improving the robustness of phononic crystal-based sensors. Imanian [7] investigated a variety of quasi-periodic structures such as Fibonacci, Thue-Morse, and Rudin-Shapiro sequences for gas sensings. These structures employed solid–liquid layered configurations, and the longitudinal wave propagation characteristics were analyzed using the transfer matrix method. However, the quality factors (Q-factors) of existing phononic crystal-based gas sensors remain limited, hindering further development. Therefore, introducing additional physical mechanisms to improve Q-factors is necessary.

Bound states in the continuum (BICs) refer to localized modes whose energy lies within the continuum spectrum yet do not radiate energy outward, theoretically exhibiting infinitely high Q-factors. Huang et al.[8] were the first to design and experimentally verify three types of acoustic BIC modes in an open acoustic resonator, achieving Q-factor breakthroughs through symmetry protection, modal interference, and mirror symmetry mechanisms to confine acoustic energy. Tong et al.[9] proposed a new paradigm of topological mechanical BICs, leveraging topological properties to revolutionize phonon control systems and provide experimental platforms for high-frequency precision measurements. Cao [10] designed an elastic metamaterial absorber based on the quasi-BIC mechanism. By employing subwavelength structures, the Friedrich–Wintgen mode was excited to achieve perfect absorption under extreme conditions. Deriy et al. [11] realized an all-solid-state BIC within a solid resonator, revealing the influence of geometric symmetry on BIC evolution and overcoming limitations of open systems. Z. Zhou et al. [12] combined quasi-BICs with orbital angular momentum and realized a "perfect chiral exceptional point" by breaking symmetry, thereby expanding the degrees of freedom for chiral manipulation.

Current studies on BICs in acoustic resonators are primarily focused on open cavity models, where the Q-factor is enhanced by adjusting the width of connected waveguides. BICs represent idealized resonant states with theoretically infinite Q-factors, but in practical engineering applications, true BICs cannot be realized due to constraints such as material losses, fabrication imperfections, and non-ideal boundary conditions. Therefore, only their approximate form—quasi-BICs (QBICs) can typically be achieved. Although theoretical and simulation studies report QBICs with Q-factors reaching as high as $10^9$ or more, experimentally obtained QBICs often exhibit much lower Q-factors, leaving substantial room for improvement. Fabry–Pérot (F–P) resonance refers to the interference enhancement effect caused by multiple reflections of waves between two reflective surfaces. In acoustics, this mechanism can enhance or suppress sound waves at specific frequencies and has been widely applied in acoustic filters[13], sensor designs [14, 15], and metamaterial structures[16, 17]. It is therefore necessary to introduce the F–P resonance mechanism to improve the localization of acoustic energy: by modifying the conventional open resonator into a semi-closed cavity that prevents material exchange while allowing selective energy transmission, the Q-factor of acoustic QBICs in experiments may be significantly enhanced.

For semi-closed structures with promising engineering applications, the balance mechanism between acoustic energy leakage and mode localization has not yet been systematically investigated. These structures typically feature finite wall thicknesses, allowing sound waves of specific frequencies to partially transmit while preventing material exchange between the inside and outside of the cavity. As a result, conventional simplified acoustic models, such as those based on rigid-wall assumptions or Helmholtz resonance, fail to accurately characterize energy leakage and modal coupling in thin-walled structures. In particular, when the cavity wall is sufficiently thin, structural deformations induced by acoustic field excitation can significantly modulate the internal resonant states, resulting in a bidirectional feedback between the sound pressure field and elastic deformation. This leads to a strong multi-physical coupling behavior. Such coupling not only affects the existence conditions and tunability of QBIC/BIC states, but also makes it difficult to accurately predict the evolution of the system's Q-factor using existing theoretical models.

In this work, we investigate how acoustic–solid coupling in semi-closed structures influences the formation conditions and control mechanisms of QBIC/BIC states, and reveal the influence of key structural parameters on the Q-factor. Based on this, we achieve high-Q QBICs and demonstrate their application in highly sensitive gas-concentration detection. Furthermore, laser Doppler vibrometry is employed to image the acoustic field and verify the sensor's stable response to varying gas concentrations.

# 2. Friedrich–wintgen QBIC behaviors based on acoustic–solid couplings

Fig. 1 shows the model of the proposed semi-closed QBIC resonant acoustic cavity. The cavity has a length of 100 mm along the *x*-direction. On both sides of the cavity, there are circular openings with a diameter of 50 mm, which are covered and sealed by 0.1 mm thick circular aluminum plates. Acoustic waves are incident from one side in the form of plane waves, which enter the cavity to excite the acoustic field inside.

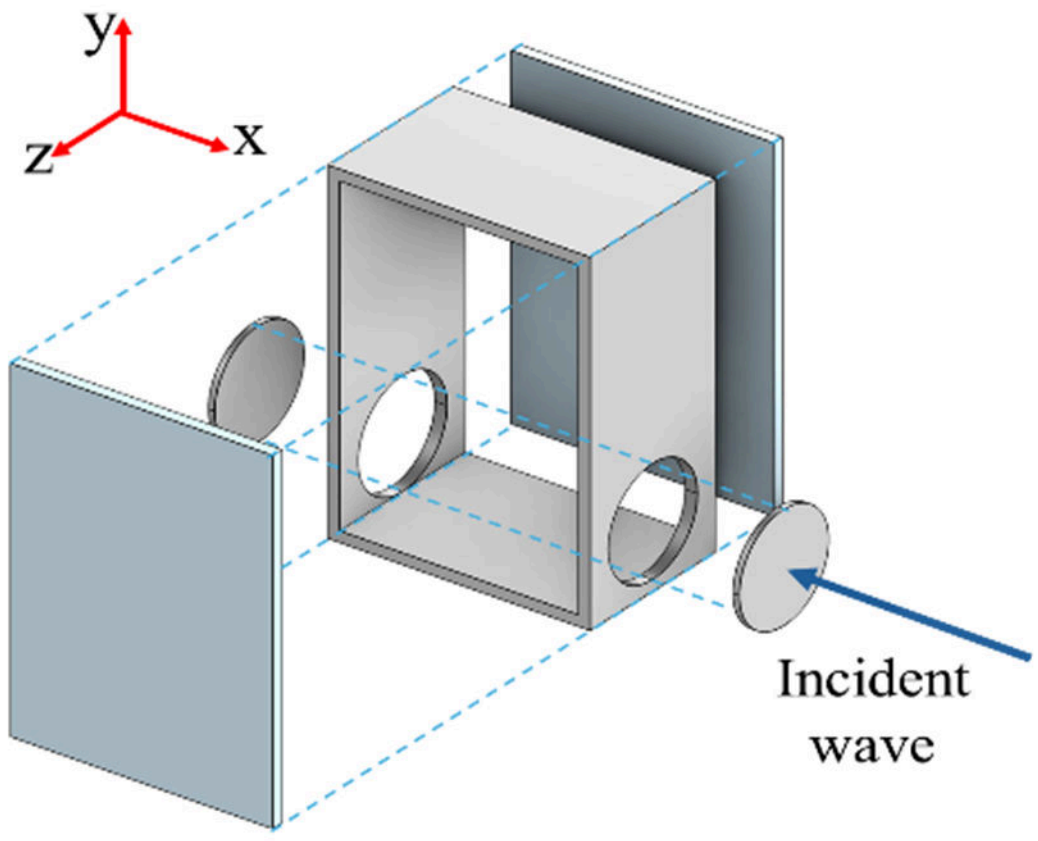

Fig. 1. Exploded view diagram of the acoustic cavity

## 2.1 Realization of Friedrich–Wintgen QBIC based on acoustic-solid couplings

To analyze the coupling between Fabry–Pérot resonances and eigenmodes in the semi-closed acoustic cavity, a two-dimensional planar acoustic model is established. Theoretical calculations based on modal analysis, transfer matrix method, and coupled mode theory are carried out to reveal the formation mechanism of the Friedrich–Wintgen QBIC.

At the eigenfrequencies, the acoustic cavity exhibits corresponding eigenmodes, and the total acoustic field inside the cavity can be expressed as a weighted superposition of all eigenmodes. The cavity considered in this section is filled with air. To obtain the eigenmodes of the rectangular cavity with impedance boundary conditions, the acoustic wave equation inside the cavity is established,

namely

$$\nabla^2 p(\mathbf{r},\omega) + k_0^2 p(\mathbf{r},\omega) = -\zeta(\mathbf{r},\omega), \tag{1}$$

$$(k_0 = \omega / c_0), \tag{2}$$

where $\zeta(r,\omega)$ represents the distribution of volumetric acoustic sources entering the cavity from the exterior, and the cavity boundary is subject to impedance boundary conditions:

$$\left[\frac{\partial p(\mathbf{r},\omega)}{\partial n} - ik_0\beta(\mathbf{r},\omega)p(\mathbf{r},\omega)\right]_s = 0, \tag{3}$$

and the eigenmodes and corresponding eigenfrequencies are defined as the nontrivial solutions to the following homogeneous problem:

$$\nabla^2\Psi_\lambda + \left(\frac{\Omega_\lambda}{c_0}\right)^2 \Psi_\lambda = 0, \tag{4}$$

$$\left[\frac{\partial \Psi_\lambda}{\partial n} - ik_0\beta(\mathbf{r},\omega)\Psi_\lambda\right]_s = 0. \tag{5}$$

The eigenfrequencies are generally complex, however the acoustic field can be expressed as a superposition of the eigenmodes $\Psi_\lambda$, namely

$$p(\mathbf{r},\omega) = \sum_{\lambda=0}^{\infty} a_\lambda \psi_\lambda(\mathbf{r},\Omega_\lambda). \tag{6}$$

Substituting the above expression into the acoustic wave equation (1) yields

$$\sum_{\lambda=0}^{\infty} a_\lambda \left[k_0^2 - \left(\frac{\Omega_\lambda}{c_0}\right)^2\right]\psi_\lambda(\mathbf{r},\Omega_\lambda) = -\zeta(\mathbf{r},\omega), \tag{7}$$

Both sides of Eq. (7) are multiplied by $\Psi_\mu$ and integrated, yielding the weight expression of the eigen mode:

$$a_\lambda = -\frac{1}{N_\lambda^2\left[k_0^2 - (\Omega_\lambda/c_0)^2\right]}\int_V \zeta(\mathbf{r},\omega)\psi_\lambda(\mathbf{r},\Omega_\lambda)\,\mathrm{d}^3\mathbf{r}, \tag{8a}$$

$$N_\lambda^2 \equiv \iint_V \psi_\lambda^2\,\mathrm{d}^3\mathbf{r}. \tag{8b}$$

Due to the impedance boundary condition of the acoustic cavity, attenuation occurs across the entire eigenmode. Therefore, the resonance frequency satisfies the following equation:

$$\Omega_\lambda = \Omega_\lambda^r + i\Omega_\lambda^i, \tag{9}$$

$$\mathrm{Re}\left[k_0^2 - \left(\frac{\Omega_\lambda}{c_0}\right)^2\right] = \frac{1}{c_0^2}\mathrm{Re}\left[\omega^2 - (\Omega_\lambda^r + i\Omega_\lambda^i)^2\right] = 0, \tag{10}$$

$$\omega_{R=}\sqrt{(\Omega_\lambda^r) - (\Omega_\lambda^i)^2} = \Omega_\lambda^r\sqrt{1 - \left(\frac{\Omega_\lambda^i}{\Omega_\lambda^r}\right)^2} \approx \Omega_\lambda^r - \frac{1}{2}\frac{(\Omega_\lambda^i)^2}{\Omega_\lambda^r}. \tag{11}$$

When the cavity walls are approximately rigid, the set of eigenmodes can be approximated as an orthogonal and complete basis, and the acoustic field can be expanded accordingly, namely

$$\begin{aligned} p(\mathbf{r},\omega)c_\lambda &\approx -\int_V \zeta(\mathbf{r},\omega)\left[\sum_{\lambda=0}^{\infty}\frac{1}{k_0^2 - k_\lambda^2 + \chi_{\lambda\lambda}}\psi_\lambda(\mathbf{r},\omega_\lambda)\psi^*{}_\lambda(\mathbf{r}',\omega_\lambda)\right]d^3\mathbf{r}' \\ &= \int_V G(\mathbf{r},\mathbf{r}',\omega)\zeta(\mathbf{r}',\omega)d^3\mathbf{r}'. \end{aligned} \tag{12}$$

Therefore, the pressure distribution in the acoustic field is closely related to the position of the acoustic source. The Green's function is defined as

$$\mathrm{G}(\mathbf{r},\mathbf{r}',\omega) \equiv \sum_{\lambda=0}^{\infty}\frac{1}{k_0^2 - k_\lambda^2 + \chi_{\lambda\lambda}}\psi_\lambda(\mathbf{r},\omega_\lambda)\psi^*{}_\lambda(\mathbf{r}',\omega_\lambda). \tag{13}$$

When standing waves are generated inside the cavity, Fabry–Perot (F-P) resonance emerges, the acoustic field can penetrate the entire structure. In this case, the transmission spectrum obtained from the transfer matrix method can be used as a known condition, namely

$$T_{total} = T_n \cdot \cdots \cdot T_2 \cdot T_1 = \begin{pmatrix} T_{11} & T_{12} \\ T_{21} & T_{22} \end{pmatrix}. \tag{14}$$

The relationship between the state vectors at the input and output ends is given by

$$\begin{pmatrix} 1 & 1 \\ \frac{1}{Z_0} & -\frac{1}{Z_0} \end{pmatrix}\begin{Bmatrix} C_i \\ C_r \end{Bmatrix} e^{i\omega t} \cdot T_{\mathrm{total}} = \begin{Bmatrix} 1 \\ \frac{1}{Z_t} \end{Bmatrix} C_t e^{i\omega t}. \tag{15}$$

The frequency-dependent acoustic pressure at the inner wall of the cavity can be obtained by multiplying the output acoustic pressure and particle velocity amplitudes by the inverse of the transfer matrix of the final layer of the medium. Mathematically, this can be expressed as:

$$T_n \begin{Bmatrix} p_n \\ u_n \end{Bmatrix}\bigg|_{x=0} = \begin{Bmatrix} 1 \\ \dfrac{1}{Z_t} \end{Bmatrix} C_t, \tag{16}$$

$$\begin{Bmatrix} p_n \\ u_n \end{Bmatrix}\bigg|_{x=0} = T_n^{-1} \begin{Bmatrix} 1 \\ \dfrac{1}{Z_t} \end{Bmatrix} C_t. \tag{17}$$

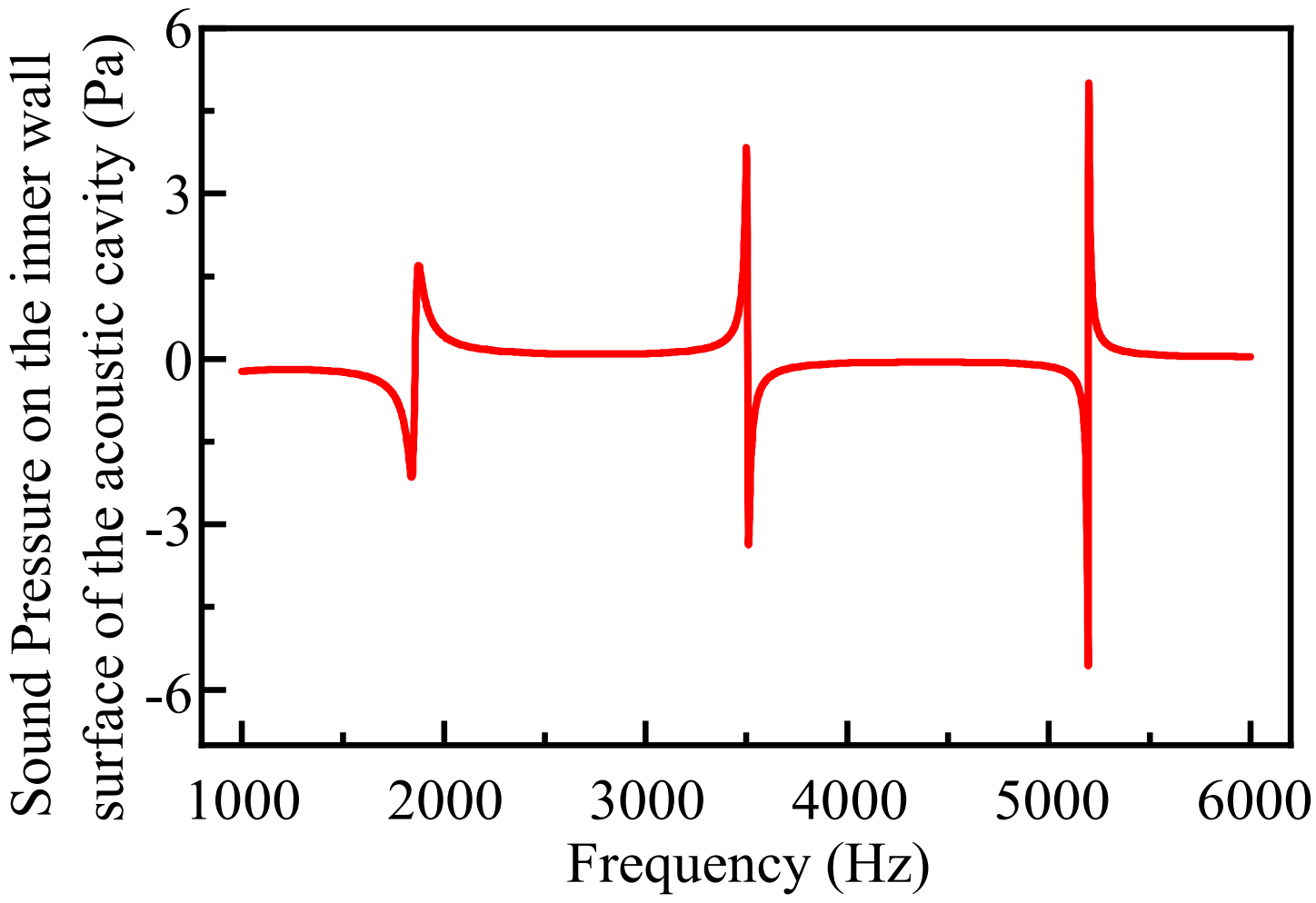

Fig. 2. Acoustic pressure spectrum on the inner wall surface of the acoustic cavity calculated via the transfer matrix method

The coupling coefficient $W$ between modes can be obtained by integrating the product of the pressure distributions of the cavity eigenmode and the Fabry–Pérot resonance mode, namely

$$W_{m,n} = \int_{-\frac{L_x}{4}-\frac{d}{2}}^{-\frac{L_x}{4}+\frac{d}{2}} \psi_{m,n}\left(x=-\frac{L_x}{2}, y\right)\phi\left(x=-\frac{L_x}{2}, y\right)dy, \tag{18}$$

where $\Psi$ $(m,n)$ denotes the pressure distribution of the corresponding cavity eigenmode, and $\Phi$ represents the acoustic pressure on the inner side of the aluminum plate associated with the Fabry–Pérot resonance. The effective Hamiltonian of the system is then constructed based on the coupling coefficients $H_{\mathrm{eff}}$.

$$H_{\mathrm{eff}} = H_R - \sum_C ik_p W_{Cp} W_{Cp}^{+}, \tag{19}$$

where $H_R$ denotes the Hamiltonian of the closed cavity, and $k_p$ is the wavenumber associated with waveguide propagation. By solving the eigenvalue problem of the Hamiltonian, the complex eigenfrequencies $\omega_0 + \gamma \mathrm{i}$ are obtained[18, 19], namely

$$T(R) = \frac{(\omega-\omega_0)^2\cos^2\phi + \gamma^2\sin^2\phi \pm 2\sin\phi\cos\phi(\omega-\omega_0)\gamma}{(\omega-\omega_0)^2+\gamma^2} \tag{20}$$

In Fig. 3, the calculated transmission-frequency curve shows good agreement with simulation results.

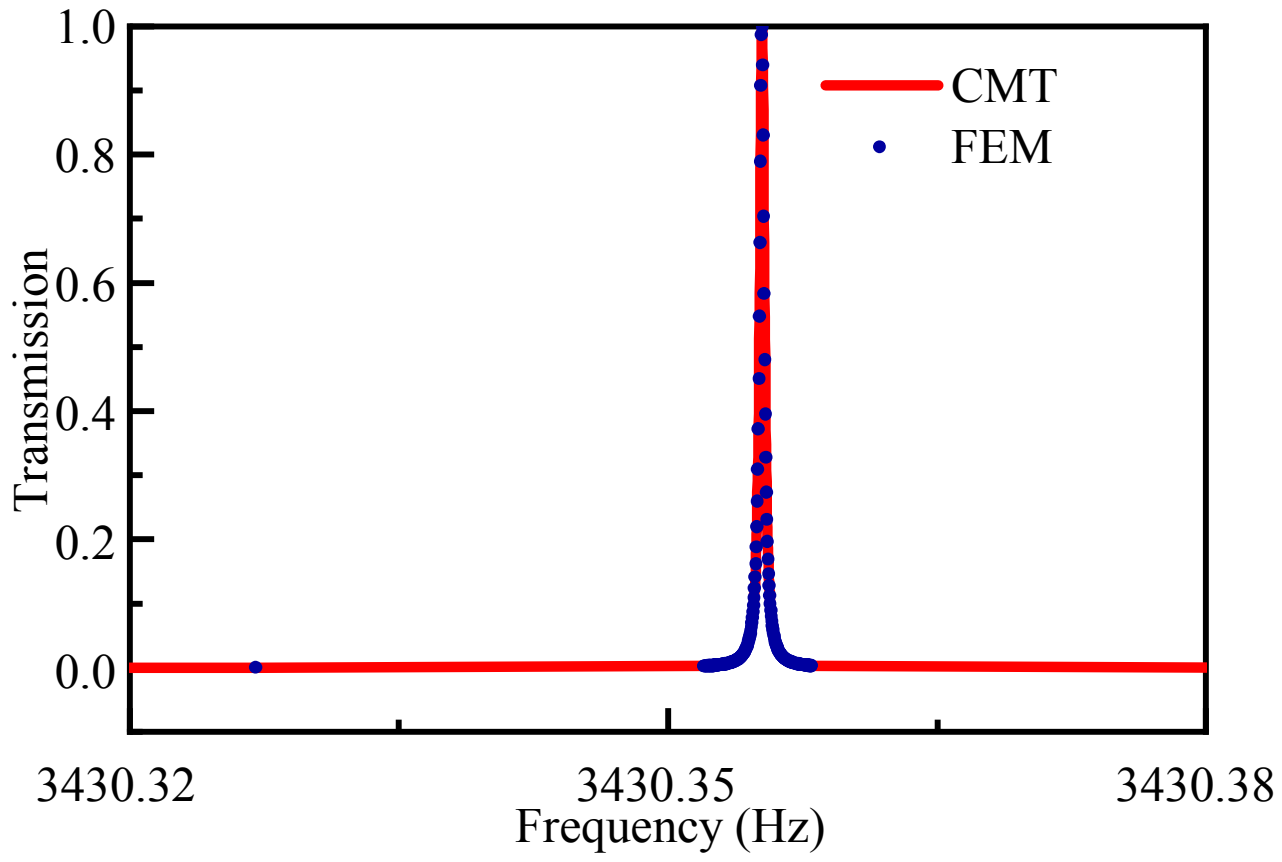

Fig. 3. Transmission coefficients calculated by the finite element method and coupled mode theory

To verify the resonance behavior of the Friedrich–Wintgen QBIC within the cavity, acoustic pressure distribution simulations were conducted by using COMSOL, the results are shown in Fig. 4. The pressure acoustics physics interface was employed to calculate the sound pressure in all gas regions, with perfectly matched layers (PMLs) added at both ends to simulate wave dissipation in two semi-infinite media. The structural mechanics physics interface was used to simulate elastic wave propagation within the aluminum layer, with the interface between the aluminum and air domains defined as an acoustic-structure boundary. A background pressure field was applied in the air domain on the left side to simulate incident plane waves propagating horizontally toward the right. Except for the aluminum plates, the rest of the structure was filled with air. Acoustic waves with an amplitude of 1 Pa were excited in the air domain at the left of the cavity. Frequency-domain calculations were performed for the proposed thin-walled acoustic cavity over the frequency range of 1 to 3.5 kHz.

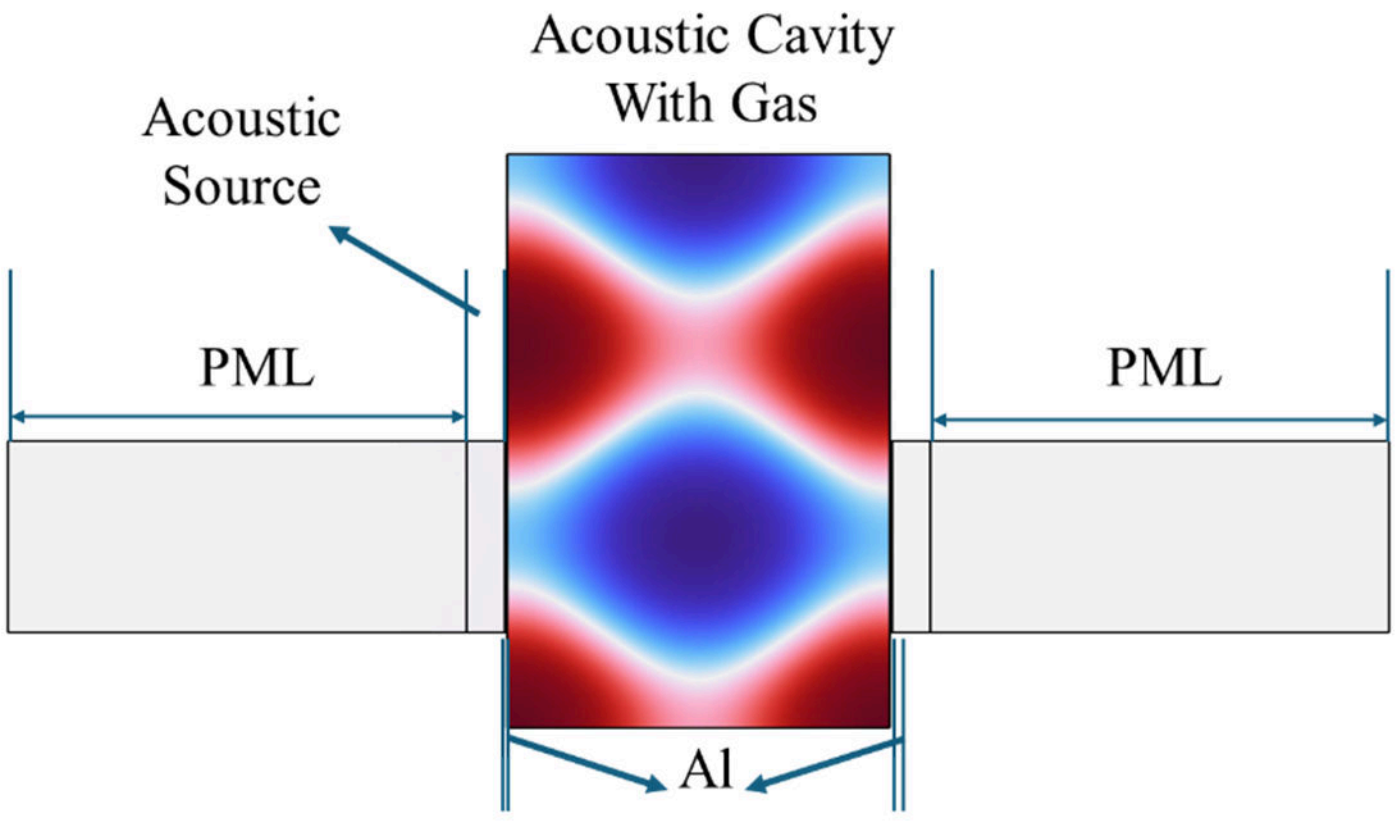

Fig. 4. Simulation configuration: Perfectly Matched Layers (PMLs) applied on both lateral boundaries for acoustic wave absorption; Acoustic source positioned at the left cavity boundary

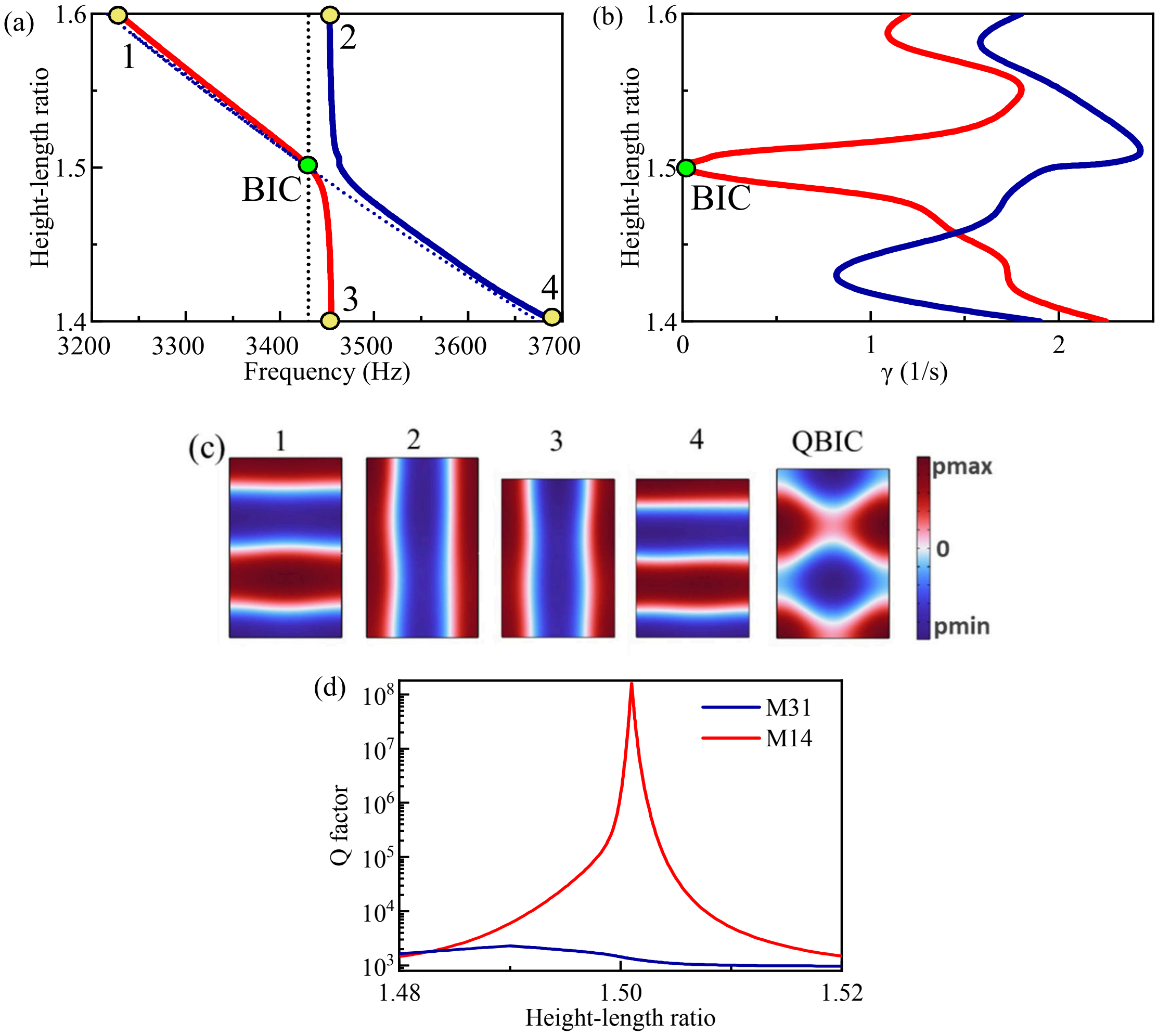

Fig. 5. (a) and (b) The condition for the appearance of QBIC , it is shown that when the height-to-length ratio equals 1.501, the optimal QBIC with a Q-factor of $1.5\times10^{8}$ emerge, (c) Mode distributions corresponding to different points, (d) Variation of Q-factors of two eigenmodes with the height-to-length ratio

Keeping the cavity length (*x*-direction) constant while varying the height (*y*-direction) to length (*x*-direction) ratio, simulations were performed on structures with height-to-length ratios ranging from 1.4 to 1.6 to determine the conditions for QBIC formation, as shown in Fig. 5. Fig. 5 (a) shows the eigenfrequencies of resonators with different dimensions. The dashed lines represent the eigenfrequencies of the closed cavity eigenmodes, while the solid lines indicate the eigenfrequencies obtained from simulations. The red lines correspond to the high-Q modes, and the blue lines correspond to the low-Q modes. $\gamma$ denotes the attenuation factor. As seen in Fig. 5 (b), when the height-to-length ratio is 1.501, the attenuation factor $\gamma=0$, indicating zero attenuation and satisfying the condition for BIC formation. The BIC point belongs to a Friedrich–Wintgen BIC, which arises from the coupling of two modes. This type of mode is characterized by weak sound pressure at the port location and strong pressure localization away from the port. Such a mode shape leads to minimal radiation of acoustic energy out of the system and energy confinement within the cavity, resulting in a localized state with a high Q-factor. Fig. 5(d) shows the variation of the Q-factors of two eigenmodes with respect to the height-to-length ratio. It is observed that as the height-to-length ratio approaches 1.501, the Q-factor of the M14 mode reaches $1.5\times10^8$, realizing a high-Q acoustic mode.

## 2.2 Influence of structural parameters on the Q-factor of Friedrich–Wintgen QBIC

### 2.2.1 Influence of port size on sensing performance

The port size is a key geometric parameter controlling the coupling strength between the external excitation and the internal acoustic modes of the cavity. Changes in port size directly affect the sound field distribution, the coupling area, and the reflection conditions, thereby significantly influencing the Q-factor. Additionally, the position and size of the ports affect the integration limits for the coupling coefficients between the two modes, ultimately influencing the sound pressure distribution. Therefore, under conditions of a fixed wall thickness and material, the port size plays an important role in determining the final Q-factor.

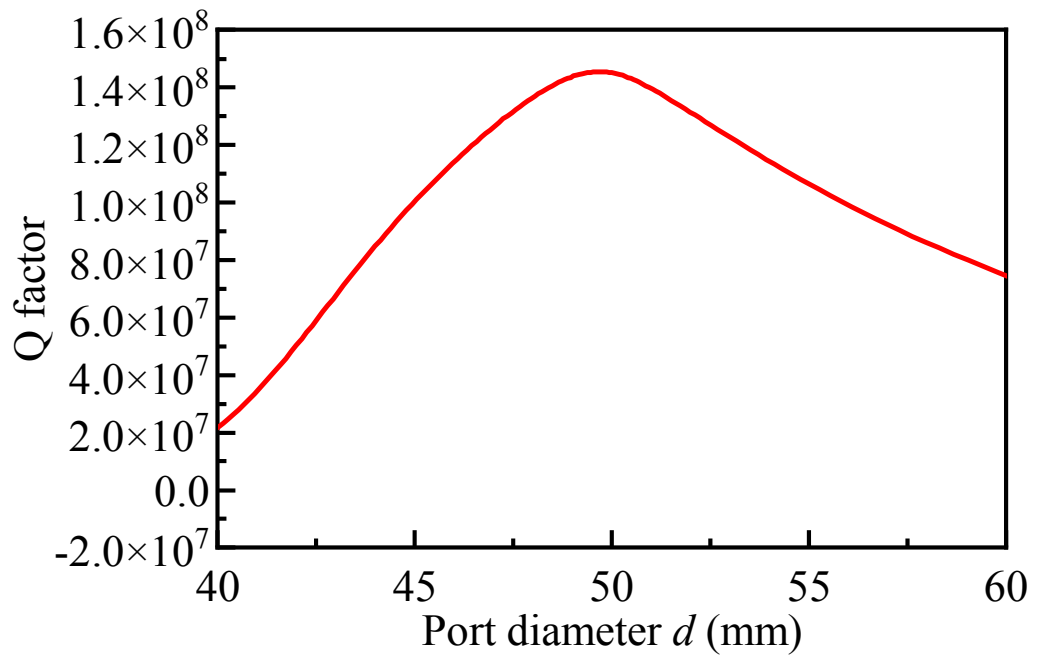

Fig. 6 Variation of the Q-factor with port diameter

The relationship between the Q-factor and port diameter is shown in Fig. 6. It can be seen that as the port diameter increases, the Q-factor rises, reaching a maximum value when the port diameter is around 50 mm. Beyond this point, further increases in port diameter cause the Q-factor to decrease. When the port size is too small, the coupling area for sound waves is limited, and the projection area of the Fabry–Pérot standing waves inside the cavity is reduced, resulting in insufficient overlap with the cavity's eigenmodes. According to coupled mode theory, the coupling coefficient is essentially the integral of the overlapping sound pressure between the two modes in the coupling region; A smaller area leads to a lower integral value, reduced modal coupling coefficient, weakened Fano interference, and difficulty of forming high-Q QBIC states. As the port size increases, the coupling region widens, allowing sound waves easier access into the cavity and expanding the Fabry–Pérot standing wave excitation area. This enhances the spatial overlap between modes, increasing the coupling coefficient and strengthening the modal interference effect, which forms a strong antisymmetric interference state, thereby increasing the Q-factor. The Q-factor reaches its maximum value near 50 mm. When the port size increases further, the system gradually departs from the "quasi-closed" condition and becomes more open, resulting in enhanced acoustic energy leakage. Calculations based on the transfer matrix method indicate that as the coupling region widens, the external sound field enters the cavity through a larger port area but also escapes more easily, causing the energy to be less effectively localized and leading to a rapid decrease in the Q-factor.

### 2.2.2 Influence of wall thickness of the cavity

In a semi-closed acoustic cavity structure, the aluminum plate not only serves as a boundary element that restricts sound wave propagation but also directly participates in the energy coupling process. By adjusting the thickness of the aluminum plate, the strength of acoustic energy exchange between the cavity and the external environment can be effectively controlled, thereby significantly impacting the system's Q-factor. The cavity wall thickness noticeably affects the resonance peak width; Theoretically, thicker walls enhance energy confinement, causing the internal cavity surfaces to experience greater acoustic pressure. However, the increased solid thickness also leads to higher acoustic losses. The following analysis focuses on the effect of wall thickness on the Q-factor in the semi-closed acoustic cavity based on the Fano resonance.

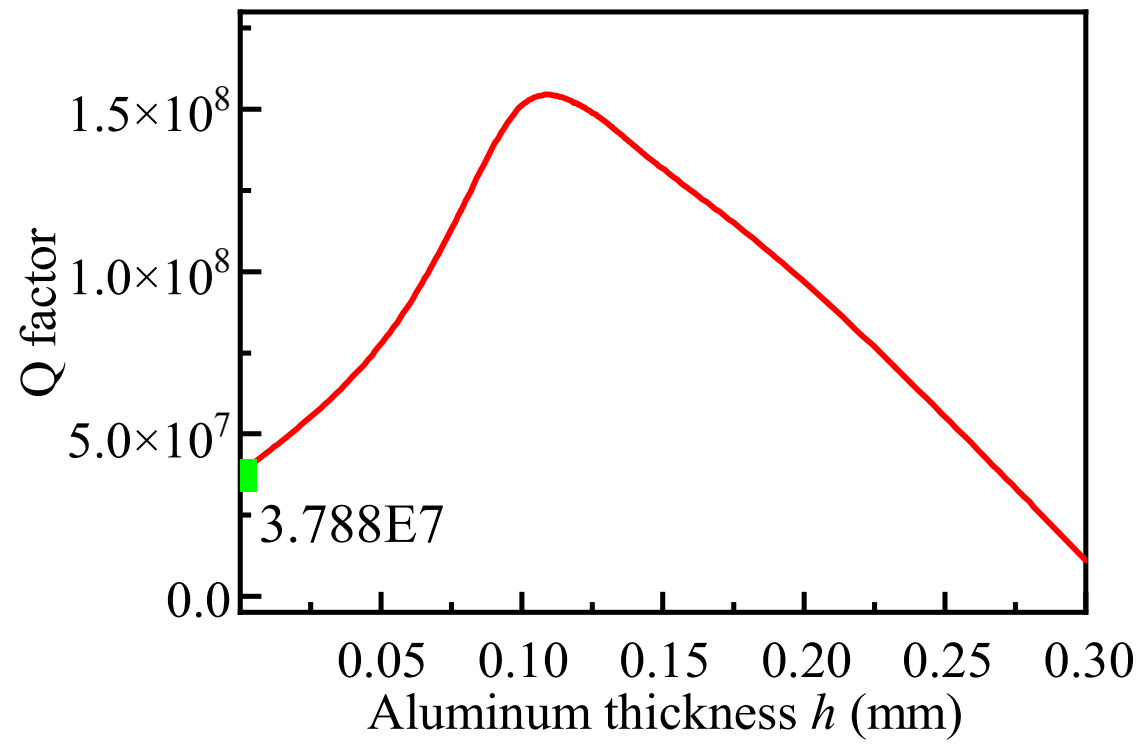

Fig. 7. Variation of the Q-factor with aluminum plate thickness

Finite element simulations were performed to calculate the Q-factors corresponding to different aluminum plate thicknesses, revealing the influence of the solid-layer thickness on the Q-factor of the Fano-resonance based gas sensor. The results are shown in Fig. 7. As the aluminum plate thickness increases, its mass effect strengthens, resulting in stronger damping on sound wave propagation. This limits the energy leakage channels and facilitates acoustic energy localization inside the cavity. Meanwhile, the reflection phase of the Fabry–Pérot structure changes with thickness, enhancing the interference matching with the cavity eigenmodes. Consequently, near a critical thickness, optimal coupling conditions are achieved, producing a typical quasi-bound state with a maximum Q-factor. As shown in Fig. 7, the Q-factor reaches its maximum at an aluminum plate thickness of 0.1 mm. However, further increases in thickness cause the plate impedance to rise significantly, greatly attenuating the coupling sound pressure at the port. Results of transfer matrix calculations indicate that the sound pressure amplitude on the inner side of the aluminum plate decreases exponentially with increasing thickness. Since the coupling coefficient in coupled mode theory is determined by the overlap integral at the port, this pressure reduction directly lowers the coupling coefficient, weakening the modal interference and thereby diminishing the QBIC state, which leads to a decline in the Q-factor. In summary, the effect of aluminum plate thickness on the Q-factor reflects a competition between two mechanisms: moderate thickness enhances energy localization, while excessive thickness reduces coupling efficiency.

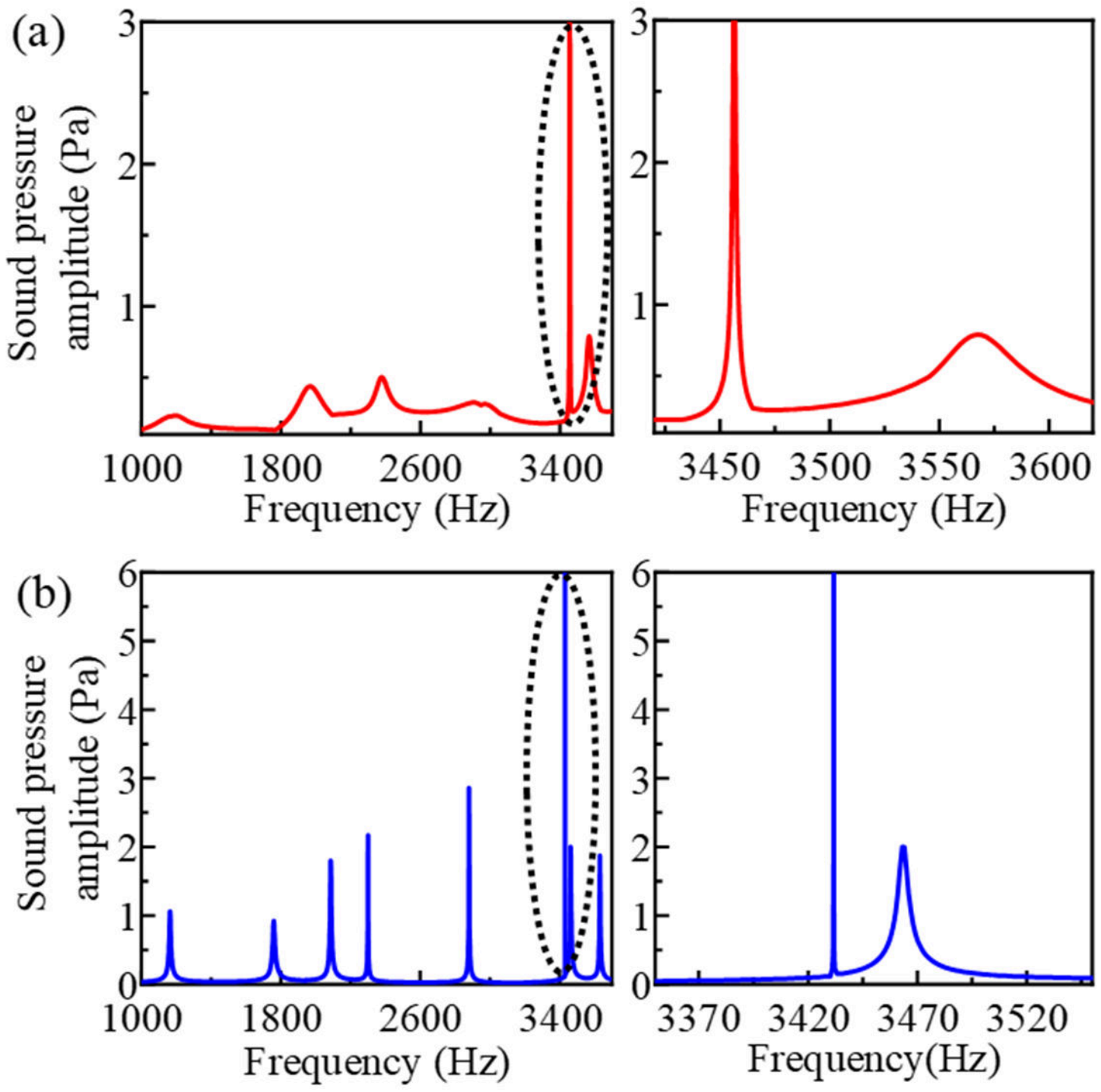

Fig. 8 Sound pressure amplitudes of QBICs in open and Fabry–Pérot resonance-type cavities with the same dimensions:(a) Sound pressure amplitude inside the cavity when all domains are filled with air (aluminum plate thickness = 0), (b) Sound pressure amplitude with an aluminum plate thickness of 0.1 mm

As shown in Fig. 8, when the aluminum plate thickness approaches zero, the system approximately degenerates into a conventional open acoustic cavity structure. In this case, acoustic energy exchanges more freely between the cavity and the external environment, resulting in significant energy leakage. Although the Fano resonance peak remains present, demonstrating that the Fano resonance mechanism exists and is stable, the Q-factor is lower compared to that when the solid layer is present. This distinction highlights an important difference between the present work and existing studies, namely compared to open resonators, the semi-closed resonator significantly sharpens the Fano resonance peak, making it more suitable for sensing applications.

### 2.2.3 Influence of solid material

The mechanical properties of the material determine the acoustic impedance and sound velocity of the solid plate, which in turn affect its ability to transmit waves, reflect waves, and form standing waves. Under the condition of unchanged geometric dimensions, the aluminum plate is replaced with various typical materials and their effects on the Q-factor of QBIC are analyzed. The results are shown in Table 1 and Fig. 9, it is found that material properties significantly influence the formation of the QBIC state.

Tab. 1 Q factors corresponding to different solid materials

| Material | Acoustic Impedance ($10^6$ kg/(m2·s)) | Q Factor ($10^7$) |
|---|---|---|
| Aluminum | 1.69 | 15.0 |
| Cast Iron | 3.50 | 2.1 |
| Copper | 4.18 | 4.2 |
| Iron | 4.50 | 5.0 |
| Steel | 4.53 | 5.3 |

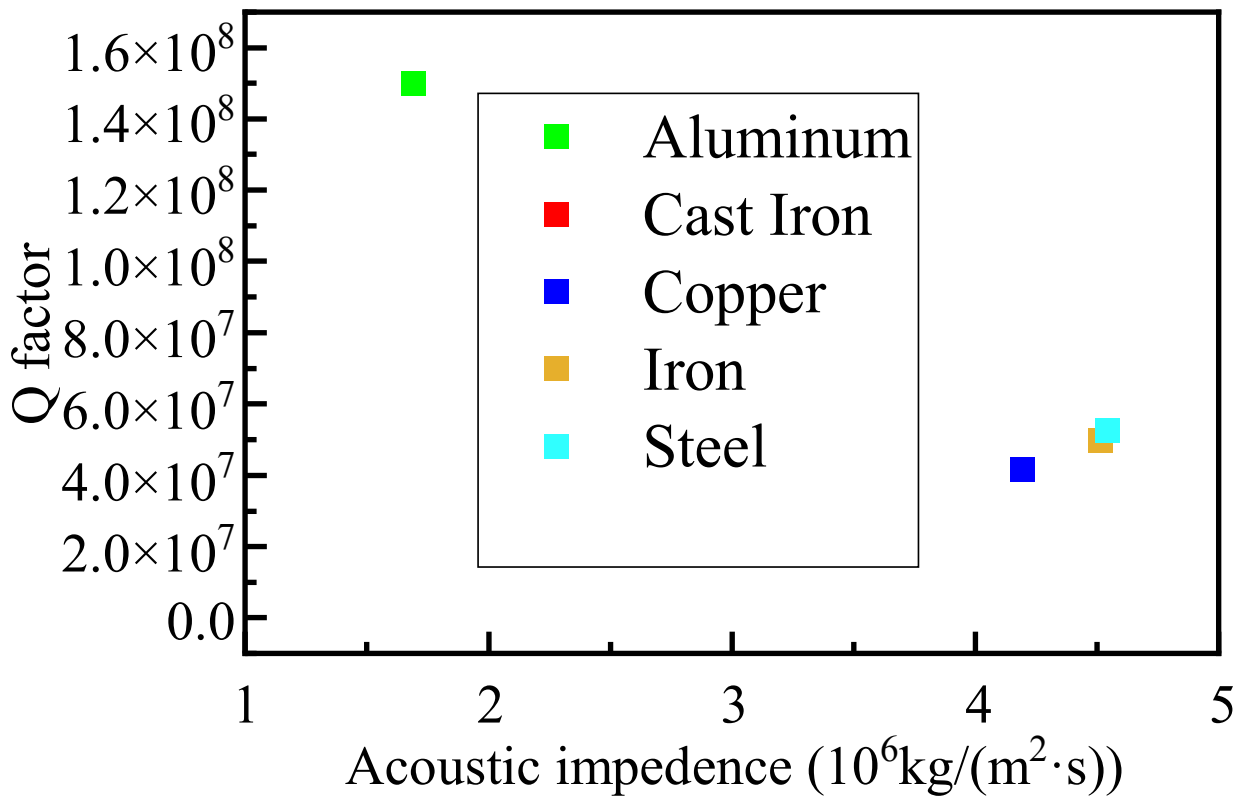

Fig. 9. Q-factors corresponding to different solid materials

Fig. 9 presents the Q factors of QBIC modes corresponding to solid layers made of different materials. First, both density and sound velocity determine the bulk acoustic impedance of a material, which directly influences the reflection and transmission coefficients of acoustic waves at the interfaces within the structure. When the impedance ratio between the solid plate and air is large, an interface with high impedance ratio reflectivity promotes the formation of Fabry-Perot standing waves, effectively enhancing acoustic energy confinement and interference. Secondly, the wave speed of different materials affects the wavelength and phase delay at a fixed plate thickness, thereby influencing the resonance conditions of the Fabry-Perot structure. For example, materials with higher wave speeds (such as steel) introduce smaller phase delays for a given plate thickness, which can lead to phase mismatch, deviating from the interference condition required to form stable QBIC modes. Conversely, moderate wave speeds (such as aluminum) match the frequency positions of cavity eigenmodes, optimizing the interference effect. Additionally, internal material losses cannot be neglected. Materials with high intrinsic losses absorb more acoustic energy internally, weakening the reflected waves, reducing coupling coefficient, and lowering the Q-factor. Low-loss metals, on the other hand, are more conducive to forming clear standing wave interference patterns and maintaining high Q-factors.

In summary, the choice of solid material should balance its acoustic impedance matching, wave speed, phase delay, and material dissipation characteristics. Comprehensive comparison shows that aluminum meets the requirements for good reflectivity, moderate wave speed, and low loss, making it be one of the optimal choices for achieving high-Q resonances in this system.

## 2.3 Experimental validation of Friedrich-Wintgen QBIC phenomena based on acousto-elastic coupling

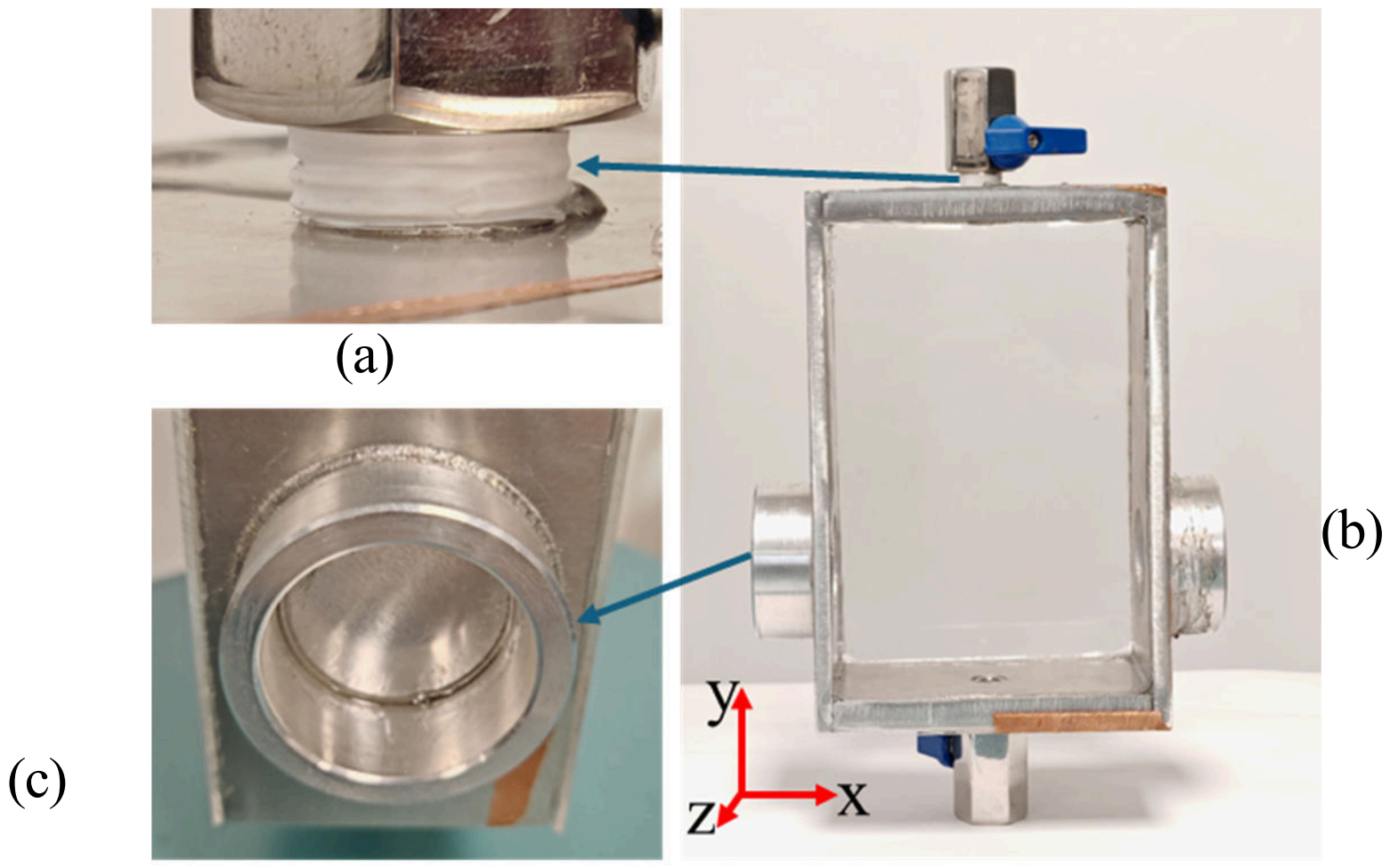

Fig. 10. Photograph of the sensing structure , (a) PTFE tape-sealed valve connection, (b) Connection to the acoustic impedance tube, (c) Acoustic cavity portion of the sensing structure

The fabricated acousto-elastic coupling QBIC device is shown in Fig. 10. The cavity has an internal length of 100 mm in the *x*-direction, an internal height of 150.1 mm in the *y*-direction, and a depth of 70 mm in the *z*-direction. The middle frame has upper and lower walls with a thickness of 10 mm and a thickness of 5 mm in the *x*-direction. Circular openings with a diameter of 40 mm are located on the left and right sides of the middle frame, with their centers positioned at 25 mm above the bottom surface. Circular aluminum plates, 0.2 mm thick and 50 mm in diameter (slightly larger than the openings to ensure a tight seal), are affixed over these openings to seal the cavity. The two openings in the z-direction on the middle frame are sealed with two transparent acrylic plates. Together, these structures form a six-sided rectangular cavity enclosing the test gas, ensuring no substance exchange with the external environment. Ball valves of specification DN8 1/4 are installed on the top and bottom of the cavity; the valve connections are wrapped with sealing tape to ensure airtightness, allowing the replacement of gases with different compositions. Around the circular aluminum plates on both sides are annular protrusions designed for connecting two sections of acoustic impedance tubes. After tightly connecting the cavity of the gas sensor to the impedance tubes on both sides, a reflective membrane is fixed at the rear position inside the transparent cavity.

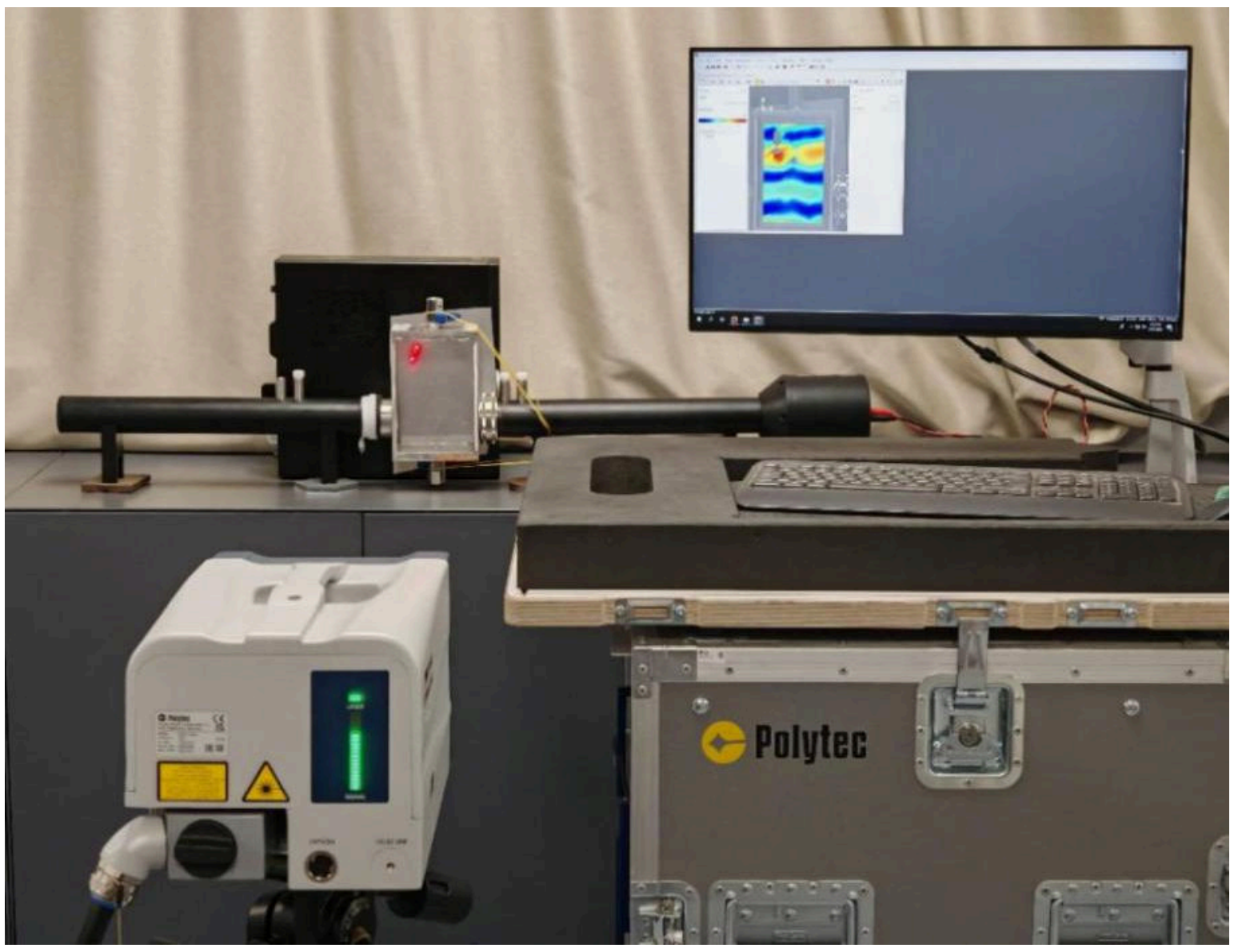

Fig. 11. Laser Doppler acoustic field measurement

The experimental testing system is shown in Fig. 11. The PSV-500 system's built-in signal generator produces a frequency-swept sinusoidal signal, which is amplified by a power amplifier and then fed into the acoustic impedance tube as an electrical excitation signal. Under an excitation, the piezoelectric speaker inside the impedance tube vibrates, generating acoustic waves that propagate as plane waves through the tube to the aluminum plate surface. Part of the acoustic wave is reflected, while the remainder enters the aluminum plate and propagates as elastic waves. After passing through two interfaces, the waves are transmitted into the gas inside the acoustic cavity and propagate therein as acoustic waves. When the external excitation frequency matches the cavity's eigenfrequency, the corresponding eigenmode is excited inside the cavity. The acoustic pressure field distribution within the cavity is measured in the frequency domain using a laser Doppler vibrometer. The variation in acoustic pressure distribution within the cavity causes changes in the refractive index of the gas inside, which in turn induces a frequency shift in the optical signal received by the vibrometer, ultimately enabling the measurement of acoustic pressure changes.

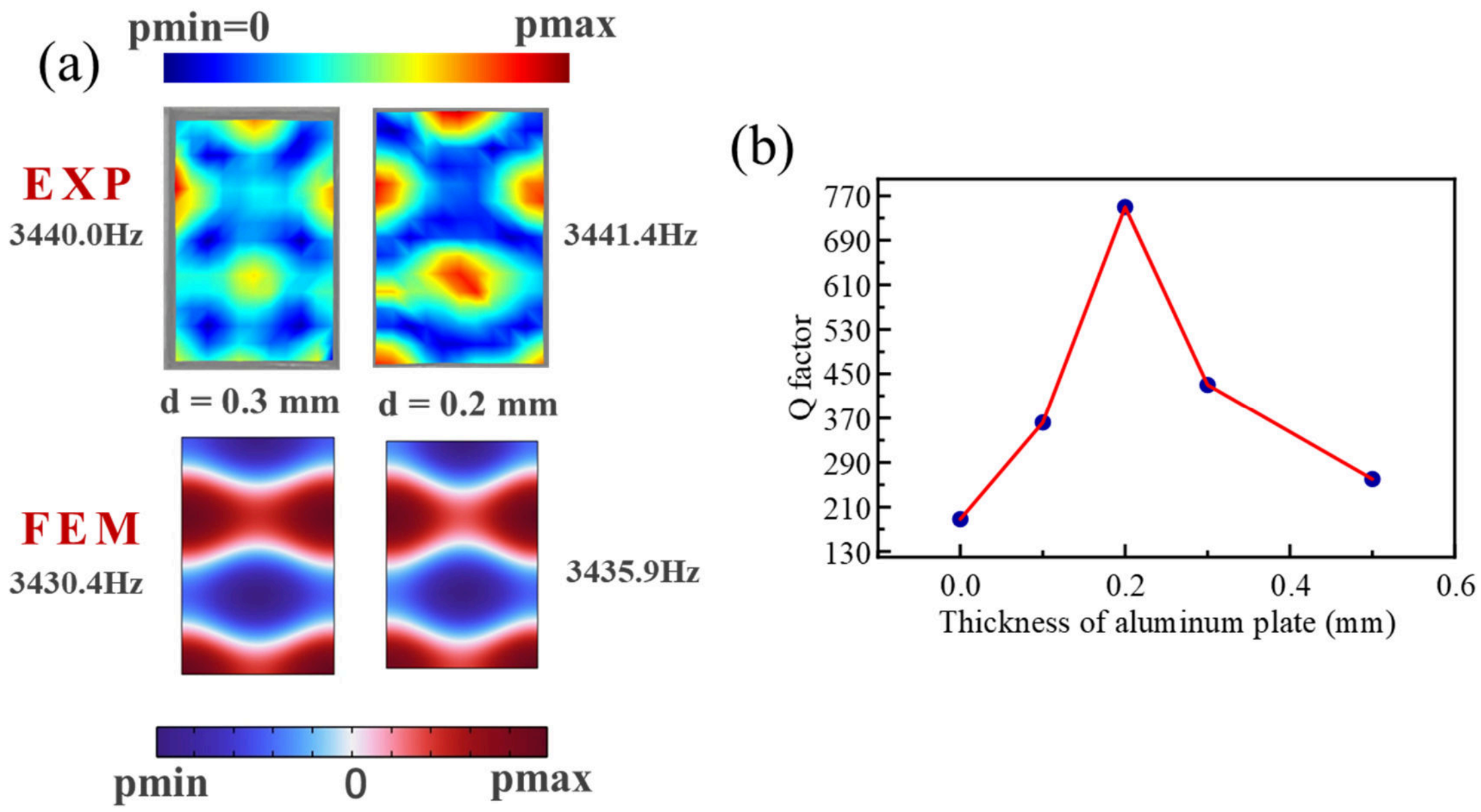

Fig. 12. Experimental results of Q factors for different thicknesses: (a) Acoustic pressure distributions corresponding to 0.3 mm and 0.2 mm aluminum plates, (b) Q factors of QBIC modes corresponding to various aluminum plate thicknesses

As shown in Fig. 12, the Q factor of the QBIC mode was measured by determining the full width at half maximum (FWHM), using $1/\sqrt{2}$ of the peak acoustic pressure as the reference pressure level. The results indicate that as the thickness of the aluminum plate increases, the Q factor of the semi-enclosed cavity first rises and then falls. When the thickness reaches 0.2 mm, the Q factor reaches maximum (750), which is obviously higher the typical value (around 400) reported in existing studies. For a fixed cavity geometry, the semi-enclosed QBIC exhibits a notably higher Q factor compared to its open counterpart. However, further increasing the aluminum plate thickness leads to a reduction in Q factor. As the transmission of acoustic energy into the cavity becomes more difficult. Consequently, the QBIC effect is weakened and eventually vanishes.

## 3. Sensing Application

When the gas composition changes, both its density and acoustic velocity also vary accordingly. This variation directly affects the acoustic impedance ratio between the gas and the solid material inside the cavity, leading to shifts in the resonance peak positions. Moreover, since the dimensions of the cavity in the $x$ and $y$ directions are fixed and designed to match either half-wavelengths or integer multiples thereof, the resonance frequency is determined by the ratio of acoustic velocity to wavelength. When the acoustic velocity of the gas changes, the resonance frequency changes proportionally in order to maintain a constant wavelength within the cavity. Based on this principle, the resonance frequency shifts and variations in the peak width caused by changes in gas composition can be used as indicators for identifying the concentration of gas components .

The Q-factor (Q) is defined as the ratio of the resonance frequency to the full width at half maximum (FWHM) of the resonance peak or the width at zero amplitude in the transmission spectrum, and it reflects the precision of the sensor.

$$Q = \frac{f_{\mathrm{r}}}{f_{\mathrm{HBW}}}, \tag{4.9}$$

where $f_r$ denotes the resonance frequency, and $f_{\mathrm{HBW}}$ is the half-power bandwidth (full width at half maximum) of the corresponding peak. A higher measured Q-factor indicates a sharper resonance peak, which enhances the frequency resolution. Sharper resonance peaks can more accurately reflect the specific conditions of the external environment and provide a pronounced response to subtle environmental changes. Therefore, employing the sharp modes generated by Fano resonance as sensing signatures can significantly enhance sensor performance. Under identical dimensional constraints, the sensing model proposed in this chapter demonstrates a substantial improvement in quality factor compared to phononic crystal sensors of the same size.

Sensitivity (S) is considered a key parameter that quantifies how the resonance mode shifts with changes in analyte concentration, and is defined by the following expression:

$$S = \frac{\Delta f}{\Delta C}, \tag{4.10}$$

where $\Delta f$ is the change in resonance frequency and $\Delta C$ is the corresponding change in analyte concentration.

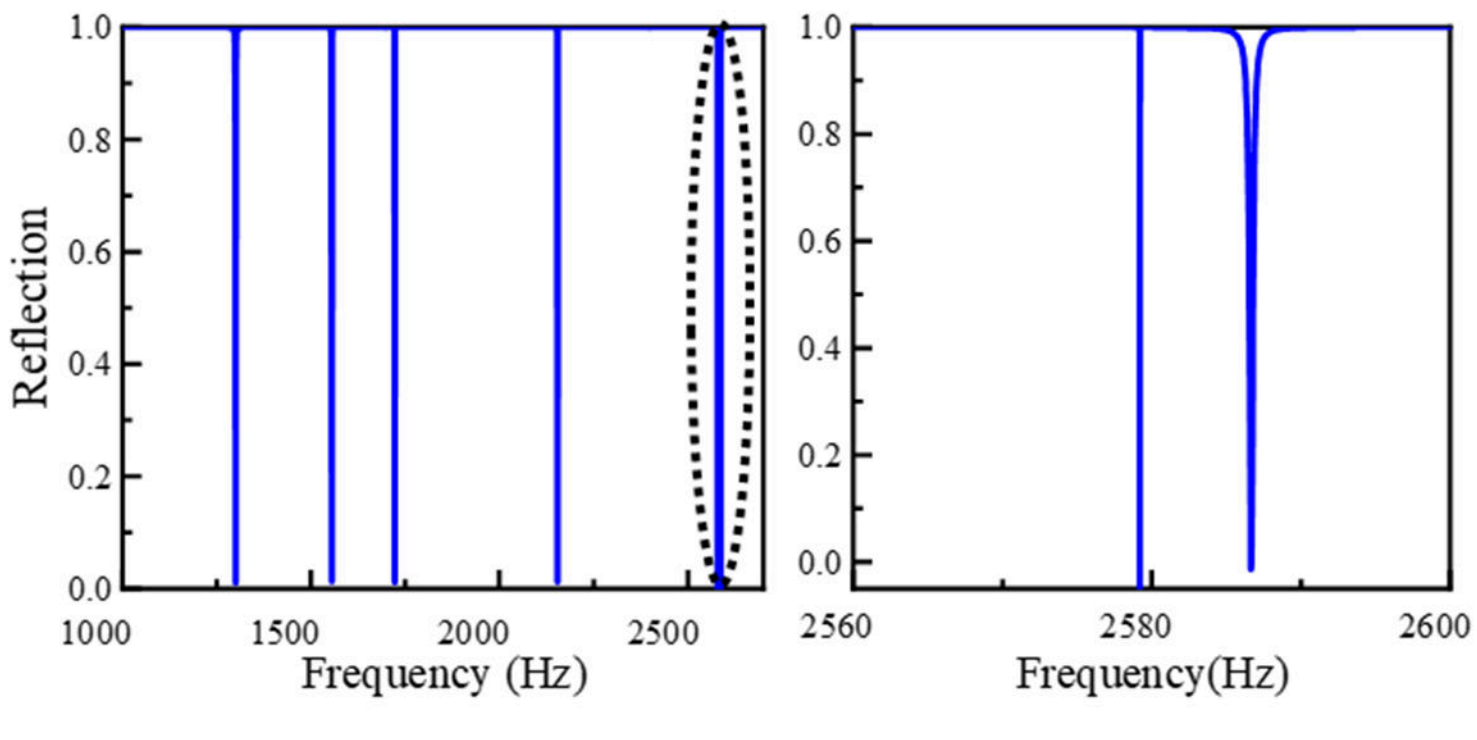

Fig. 13. Reflection spectrum of the acoustic cavity when filled with carbon dioxide

To investigate sensing performances of the proposed semi-closed acoustic cavity, we first consider the detection of pure carbon dioxide ($CO_2$). In the simulation, the gas in the cavity is set to CO by modifying the density and acoustic velocity accordingly. A frequency sweep analysis is conducted after applying the acoustic excitation. The results are shown in Fig. 13. Subsequently, the gas within the cavity and in the two semi-infinite external domains is replaced with air–CO mixtures of varying concentrations, and the sensor's frequency responses are analyzed. The results show that, after replacing the gas, multiple resonance peaks appear near specific frequencies. The acoustic pressure distributions at these resonant frequencies are consistent with those observed in air, indicating that same eigenmodes are excited. The key difference is that the frequencies at which these eigenmodes appear shift depending on the gas composition.

Taking the M31 mode as an example, when the horizontal length ($x$-direction) of the cavity matches a full wavelength, the M31 mode is excited inside the cavity. For air, this mode occurs at approximately 3430 Hz, while for CO, it occurs around 2580 Hz. Most importantly, the sharp resonance peak caused by Fano interference remains stably present even after the gas is changed.

Finite element simulations were also conducted for mixtures of air and CO with different concentrations. The acoustic velocity of these mixtures were calculated using a thermodynamic mixing model, and the results are listed in Tab. 2.

Tab. 2 Variation of acoustic velocity and gas density with CO concentration in gas mixtures

| CO Concentration (%) | Density (kg/m$^3$) | Acoustic Velocity (m/s) | QBIC Frequency (Hz) |
|---|---|---|---|
| 20 | 1.362 | 319 | 3190.4 |
| 40 | 1.494 | 300 | 3006.4 |
| 60 | 1.626 | 284 | 2846.6 |
| 80 | 1.758 | 270 | 2700.5 |
| 80.01 | 1.75807 | 269.99 | 2700.4 |
| 81 | 1.765 | 269 | 2685.1 |

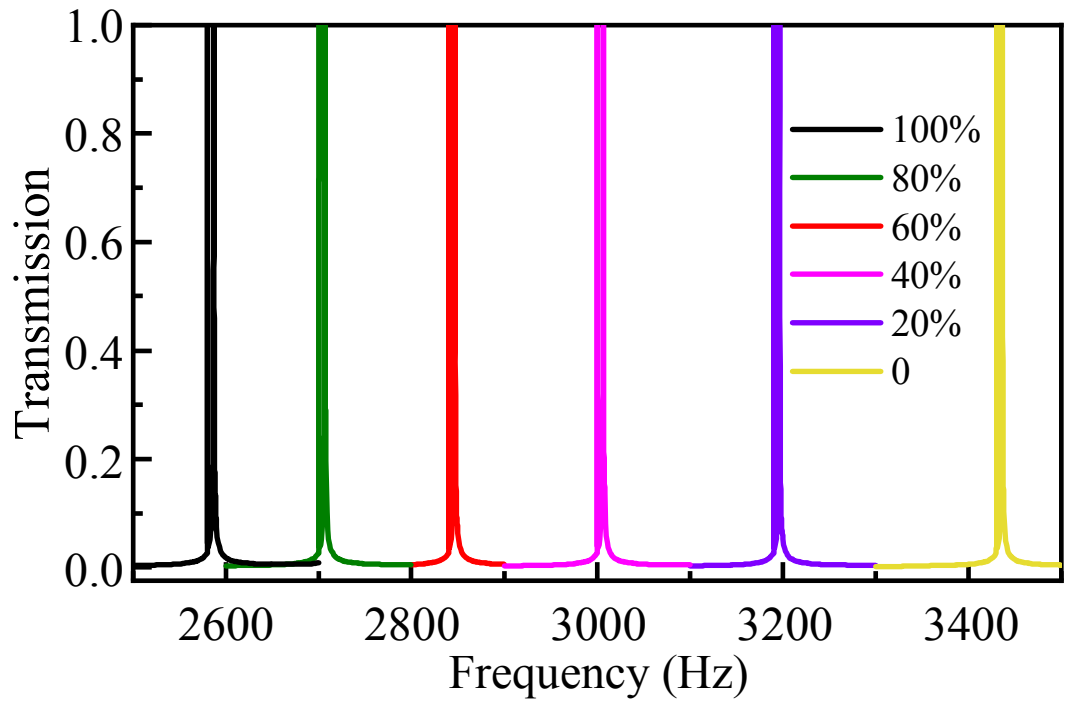

Fig. 14 Fano resonance frequencies and corresponding sound pressure amplitudes as functions of CO concentration

QBIC simulations were conducted for gas mixtures with different $CO_2$ concentrations. The solid-layer thickness in the acoustic cavity was set to 0.1 mm, and the sound source was placed at the bottom of the cavity, aligned with its lowest point, in order to excite all possible eigenmodes. The results are shown in Fig. 14. It is observed that stable Fano resonances appear for each gas composition tested, and the corresponding resonance frequencies shift significantly with changes in CO concentration. Specifically, for every 1% change in gas composition, the QBIC resonance peak shifts by approximately 10 Hz, indicating a sensitivity of 10 Hz per percent.

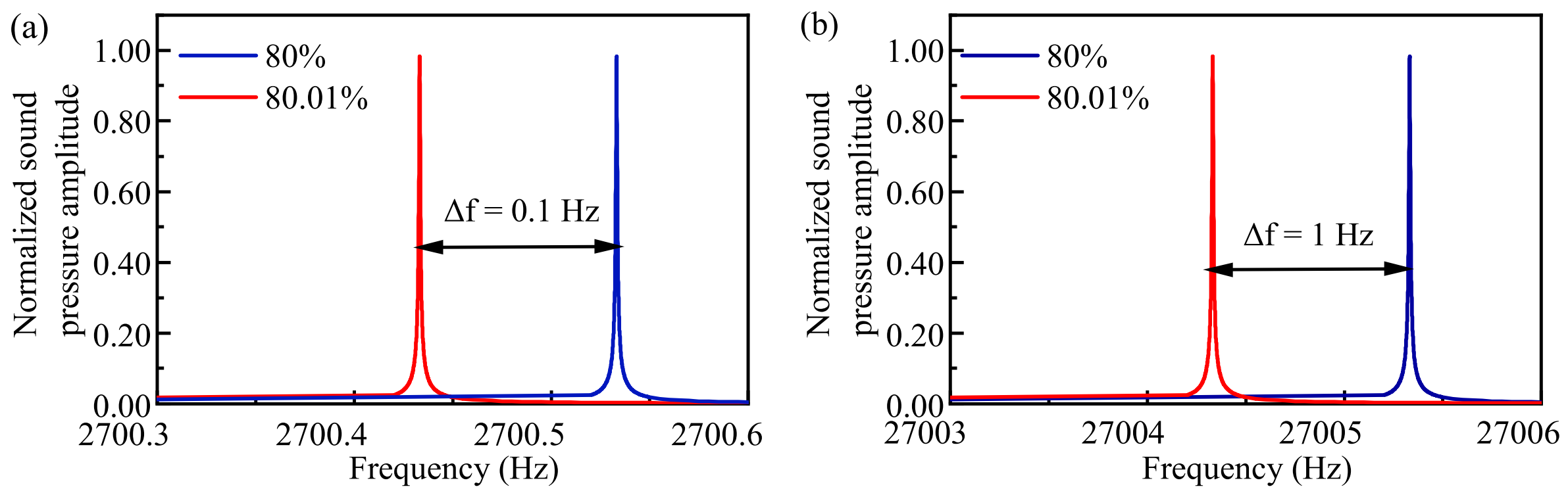

Fig. 15. Fano resonance frequencies at CO concentrations of 80% and 80.01%: (a) Cavity length = 10 cm, (b) Cavity length = 1 cm

Fig. 15 shows the frequency response of the corner-point sound pressure when the $CO_2$ is 80% and 80.01%. A clear frequency shift in the resonance peak is observed, demonstrating that the proposed sensing method offers high resolution and is capable of detecting extremely subtle changes in gas concentration. This resolution can be further enhanced by reducing the overall dimensions of the cavity.

In addition to air, the target gases were prepared by mixing high-purity gases (99.9% purity) of oxygen, nitrogen, and carbon dioxide. Other components in air, such as noble gases, were neglected. Air was approximated as a mixture of oxygen and nitrogen in a ratio of 1:4, and mixtures of air and $CO_2$ were prepared by varying concentrations of CO . Considering that oxygen and $CO_2$ are denser than air while nitrogen has a similar density, the gas mixtures were introduced into the sensor cavity and sealed using the upward method.

## 4. Experimental results of gas concentration detection

First, pure air was tested. After applying external excitation, an acoustic pressure scan of the entire cavity was conducted. The results are shown in Fig. 16 (a). At an excitation frequency of 3424 Hz, the M14 eigenmode was observed inside the cavity. Near 3440 Hz, close to the M14 mode frequency, the M31 mode was also detected. The measured acoustic pressure distributions and frequencies showed good agreement with simulation results. It should be noted that a colorless defect appears in the M14 pressure map due to anomalous measurement data, which will be discussed in detail in the next section; this defect does not affect the overall distribution pattern. Additionally, the edges of the cavity are not fully displayed because the laser source is a point beam. When expanding the laser scanning area toward the cavity edges, the laser irradiates the aluminum frame instead of the gas inside the cavity, thus detecting vibrations of the aluminum frame rather than the acoustic pressure of the gas. This issue can be effectively mitigated by increasing the distance between the cavity and the laser probe, which reduces the angle between the laser beam and the *z*-axis.

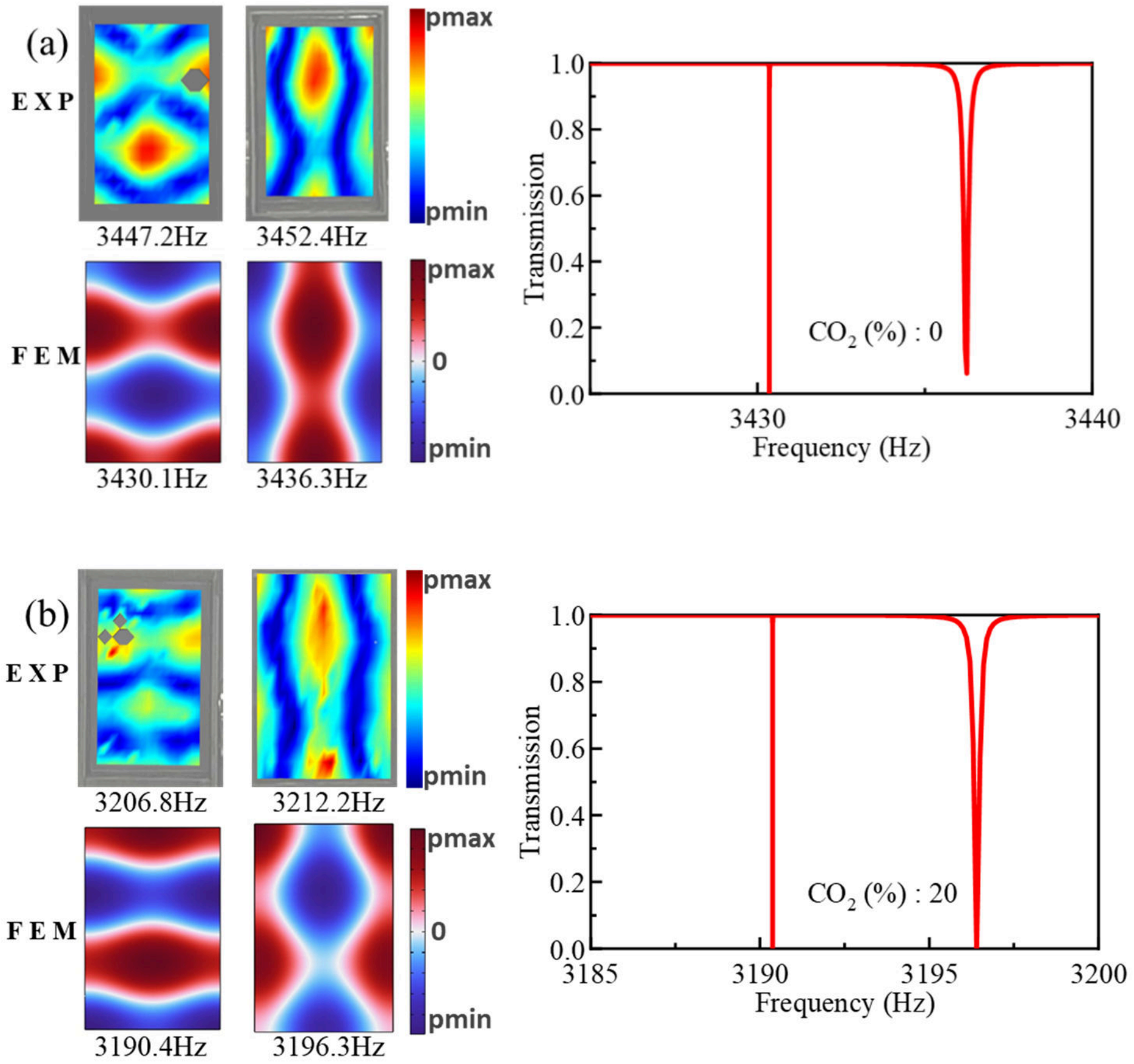

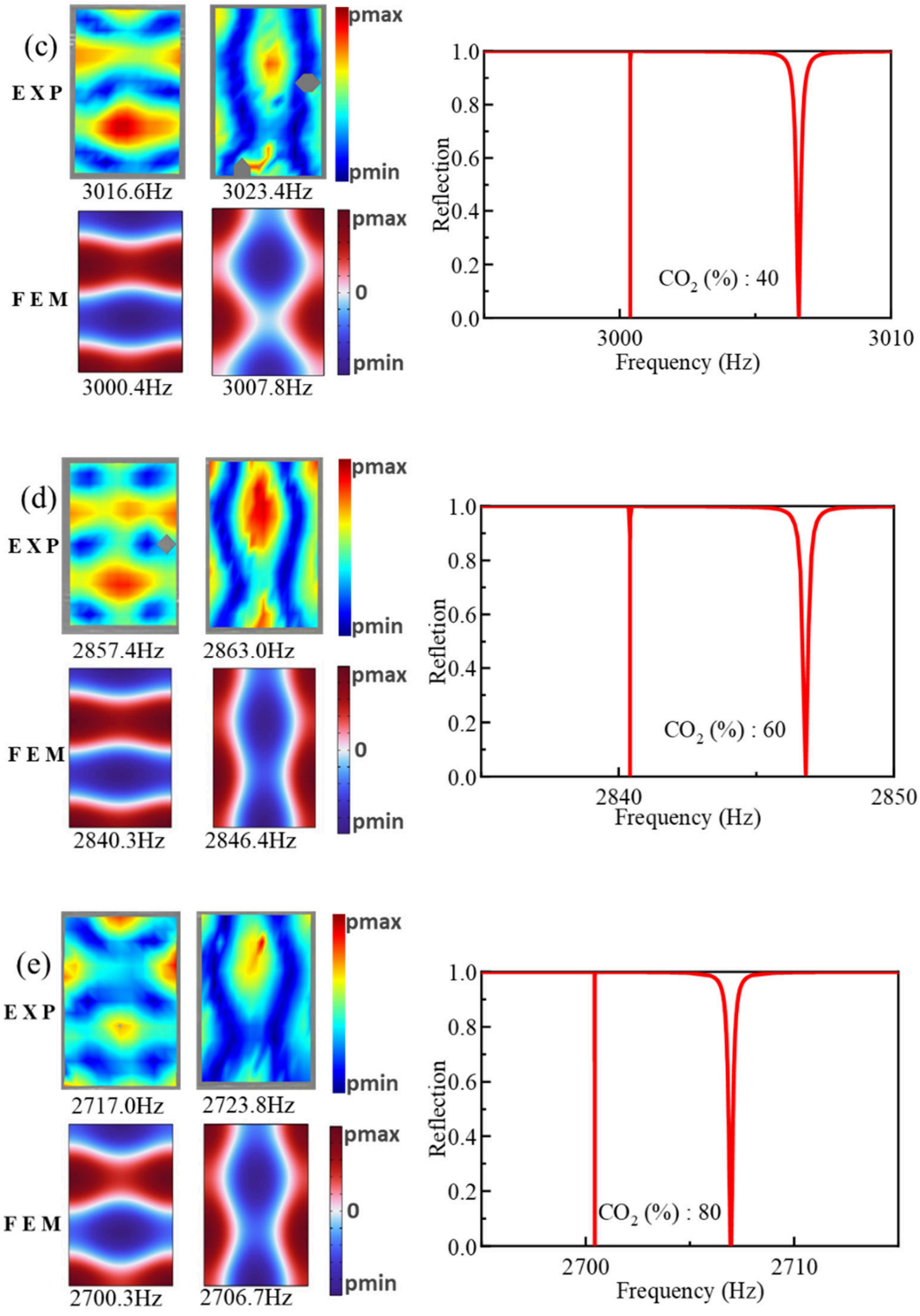

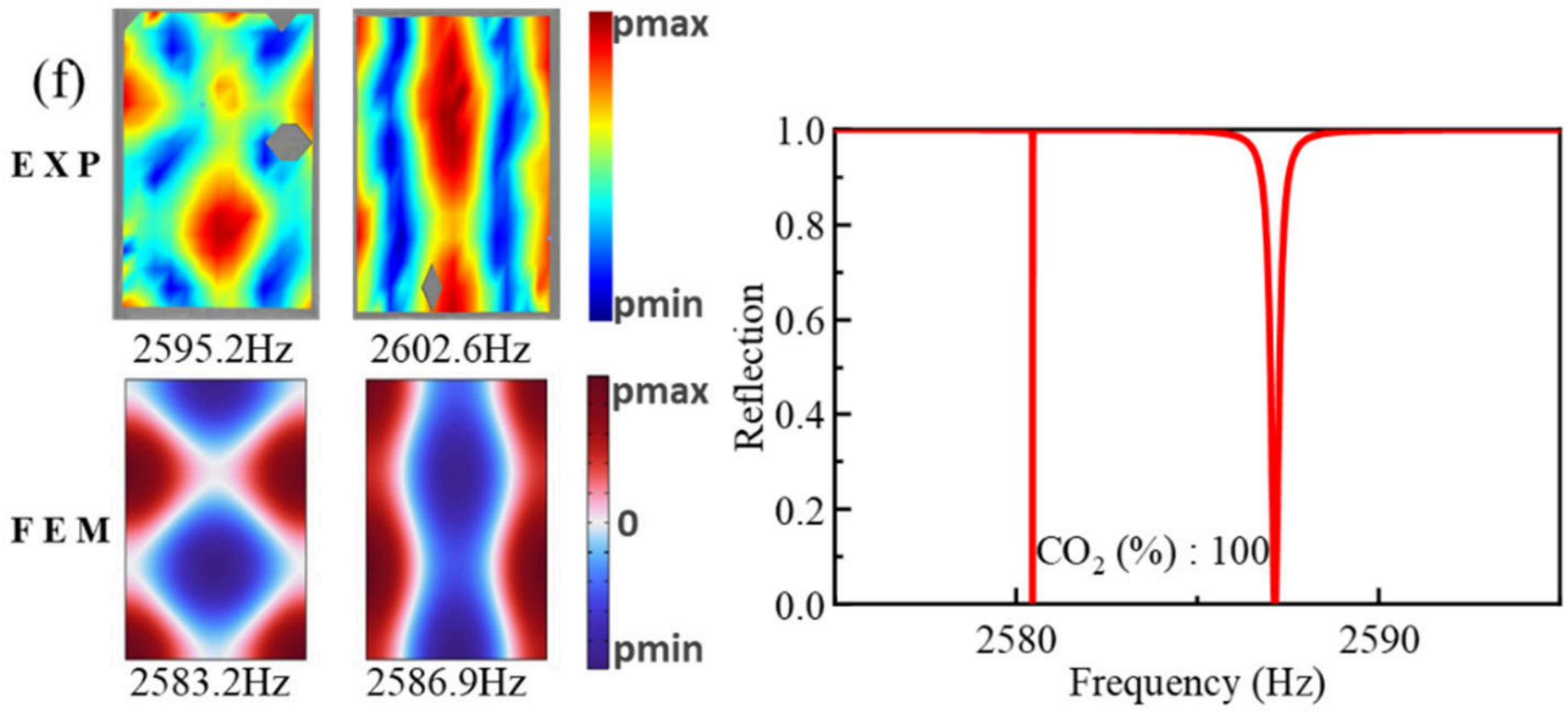

Fig. 16. Measured (top left) and simulated (bottom left) acoustic field distributions using laser Doppler vibrometry, and simulated reflection spectra (right) at a source height of 25 mm. The corresponding CO concentrations are: (a) 0%, (b) 20%, (c) 40%, (d) 60%, (e) 80%, and (f) 100%

After changing the gas composition, the eigenfrequencies and corresponding eigenmodes inside the acoustic cavity were re-measured following the original procedure. The results show that although the eigenfrequencies shift due to the altered gas composition, the spatial distribution patterns of the acoustic pressure remain unchanged, indicating that only the modal frequencies are displaced. Further tests were conducted on air–$CO_2$ mixtures with CO concentrations of 20%, 40%, 60%, 80%, and 100%, examining the frequency shifts and acoustic pressure distributions of the M31 and M14 modes. The results are presented in Fig. 16(b–f). These results demonstrate that the M31 and M14 modes persist after changing the gas composition, and their corresponding frequencies exhibit significant shifts compared to those in pure air. Experimental results show good agreement with simulation results.

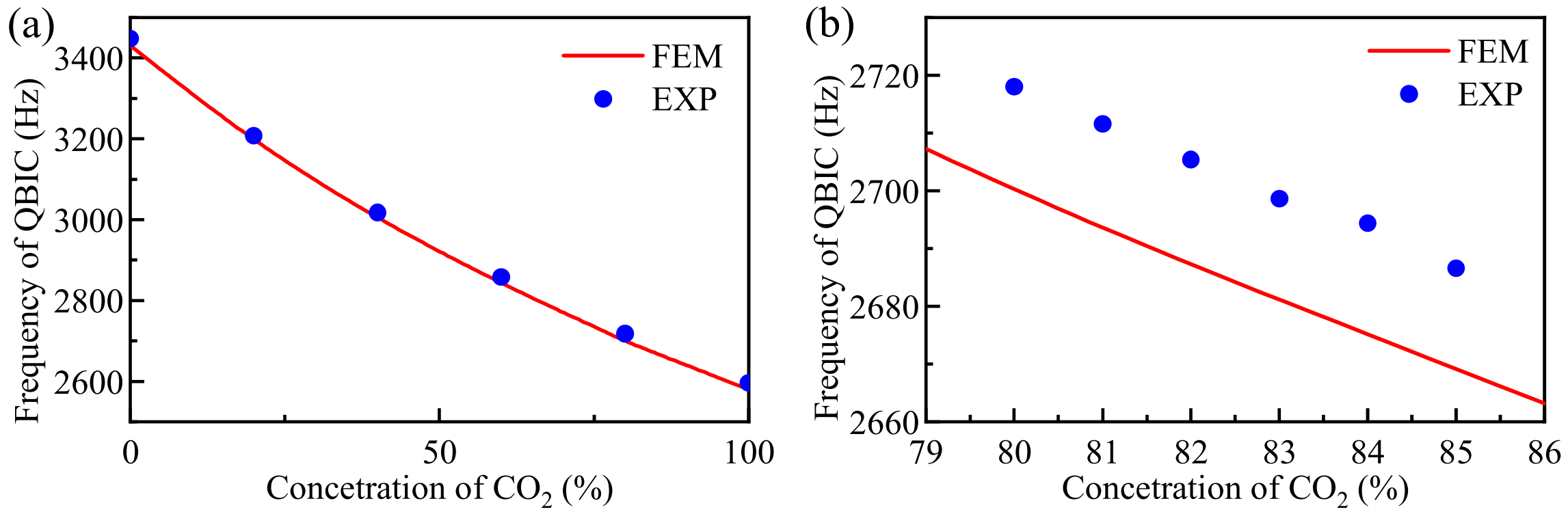

Fig. 17 Comparison of the fundamental frequencies of the experiment with different concentrations of CO and the simulation results, (a) Concentration range: 0% to 100%; (b) Concentration range: 79% to 86%

Around a $CO_2$ concentration of approximately 80%, a change of just 1% in CO composition still induces a noticeable frequency shift of 8 Hz in the corresponding eigenmode, as shown in Fig. 17. These results demonstrate that the sensing device exhibits high sensitivity to the gas concentration; even minor changes in gas composition cause stable alterations in the acoustic pressure distribution within the cavity. A consistent discrepancy of about 15 Hz between experimental and simulation results is observed, which is attributed to deviations in the measured temperature and processing errors in dimensions.

# 5. Conclusion

This work explores the formation mechanism and methodology of high-Q Friedrich–Wintgen QBIC behavior based on acoustic–solid coupling in a quasi-closed system, and investigates its application in gas sensings. First, a quasi-closed resonator model coupling elastic waves and acoustic waves was proposed. Then, the formation conditions and influencing factors of F-W BIC behavior were elucidated through coupled mode theory and finite element simulations, followed by experimental validation. The results demonstrate that Q-factors of F-W BICs realized via acoustic–solid coupling are significantly higher than those of open systems. Based on this, a gas concentration sensing technique utilizing acoustic–solid coupled F-W BIC behavior was developed, providing an important foundation for high-quality gas sensing. A gas sensor based on the acoustic–solid coupled F-W BIC was fabricated and tested. Using laser Doppler vibrometry to measure the acoustic pressure distribution and resonance frequencies, it is shown that the sensor has a significant response to varying gas concentrations, thereby verifying the feasibility and reliability of this novel gas sensing method.